# Analysis and Simulation of Delay and Buffer Requirements of Satellite-ATM Networks for TCP/IP Traffic

Rohit Goyal[1], Sastri Kota[2], Raj Jain[1], Sonia Fahmy[1], Bobby Vandalore[1], Jerry Kallaus[2]

1. Department of Computer and Information Science, The Ohio State University, 2015 Neil Ave, DL395, Columbus, OH 43210, Phone: 614-688-4482, Fax: 614-292-2911, Email: goyal@cis.ohio-state.edu

2. Lockheed Martin Telecommunications, 1272 Borregas Avenus, Bldg B/551 O/GB-70, Sunnyvale, CA 94089, Phone: 408-543-3140, Fax: 408-543-3104, Email: sastri.kota@lmco.com

## Abstract

In this paper we present a model to study the end-to-end delay performance of a satellite-ATM network. We describe a satellite-ATM network architecture. The architecture presents a trade-off between the on-board switching/processing features and the complexity of the satellite communication systems. The end-to-end delay of a connection passing through a satellite constellation consists of the transmission delay, the uplink and downlink ground terminal-satellite propagation delay, the inter-satellite link delays, the on-board switching, processing and buffering delays. In a broadband satellite network, the propagation and the buffering delays have the most impact on the overall delay. We present an analysis of the propagation and buffering delay components for GEO and LEO systems. We model LEO constellations as satellites evenly spaced in circular orbits around the earth. A simple routing algorithm for LEO systems calculates locally optimal paths for the end-to-end connection. This is used to calculate the end-to-end propagation delays for LEO networks. We present a simulation model to calculate the buffering delay for TCP/IP traffic over ATM ABR and UBR service categories. We apply this model to calculate total end-to-end delays for TCP/IP over satellite-ATM networks.

## 1 Introduction

ATM technology is expected to provide quality of service based networks that support voice, video and data applications. ATM was originally designed for fiber based terrestrial networks that exhibit low latencies and low error rates. With the widespread availability of multimedia technology, and an increasing demand for electronic connectivity across the world, satellite networks will play an indispensable role in the deployment of global networks. Ka-band satellites using the gigahertz frequency spectrum can reach user terminals across most of the populated world. As a result, ATM based satellite networks can be effectively used to provide real time as well as non-real time communications services to remote areas.

Satellite communications technology offers a number of advantages over traditional terrestrial point-to-point networks [AKYL97]. These include,

1. wide geographic coverage including interconnection of "ATM islands",
2. multipoint to multipoint communications facilitated by the inherent broadcasting ability of satellites,

3. bandwidth on demand, or Demand Assignment Multiple Access (DAMA) capabilities, and
4. an alternative to fiber optic networks for disaster recovery options.

However, satellite systems have several inherent constraints. The resources of the satellite communication network, especially the satellite and the earth station are expensive and typically have low redundancy. These must be robust and be used efficiently. Also, satellite systems use a Time Division Multiplexed (TDM) physical layer, where individual earth stations can transmit frames during fixed time slots. The cell based ATM layer must be mapped onto the frame based satellite layer. This involves the use of efficient bandwidth allocation strategies for Demand Assignment Multiple Access (DAMA) based media access techniques.

Current and proposed satellite communications networks use low earth orbit (LEO) constellations as well as geosynchronous (GEO) satellite systems. GEO satellites have a high propagation delay but a few satellites are enough to provide connectivity across the globe. LEO satellites have lower propagation delays due to their lower altitudes, but many satellites are needed to provide global service. While LEO systems have lower propagation delay, they exhibit higher delay variation due to connection handovers and other factors related to orbital dynamics [IQTC97]. The effects of the propagation delays are further intensified by the buffering delays that could be of the order of the propagation delays especially for best effort TCP/IP traffic. The large delays in GEOs, and delay variations in LEOs, affect both real time and non-real time applications. Many real time applications are sensitive to the large delay experienced in GEO systems, as well as to the delay variation experienced in LEO systems. In an acknowledgment and timeout based congestion control mechanism (like TCP), performance is inherently related to the delay-bandwidth product of the connection. Moreover, TCP Round Trip Time (RTT) measurements are sensitive to delay variations that may cause false timeouts and retransmissions. As a result, the congestion control issues for broadband satellite networks are somewhat different from those of low latency terrestrial networks. Both interoperability, as well as performance issues must be addressed before data, voice and video services can be provided over a Satellite-ATM network.

In this paper, we present a model for analyzing the delay performance of LEO and GEO satellite systems. We present an overview of our satellite-ATM network architecture. This model applies both to LEO and GEO systems. We describe the various components of the delay experienced by the cells of a connection over the satellite network. The two most important delay components are propagation and buffering delays. We present a model for calculating the propagation delay in a satellite network. We provide values for the delays experienced by connections traversing sample LEO constellations. We describe a simulation model to compute the buffer requirements of a satellite-ATM network for TCP/IP file transfer traffic. This analysis, performed for TCP/IP over ABR and UBR service categories, provides an estimate of the buffering delay experienced by a TCP/IP connection. A case study of the total delay experienced by a TCP connection over GEO and LEO systems concludes this paper.

## 2 Satellite-ATM Network Architecture

In this section, we briefly overview the basic architecture of a Satellite-ATM network. We first present a brief overview of the QoS guarantees in ATM networks. This gives the reader an idea of the kinds of guarantees that are expected of a satellite-ATM network. We then describe the various components of the architecture and overview their functionality.

## 2.1 Quality of Service in ATM Networks

ATM networks carry traffic from multiple service categories, and support Quality of Service (QoS) requirements for each service category. The ATM-Forum Traffic Management Specification 4.0 [TM4096] defines five service categories for ATM networks. Each service category is defined using a traffic contract and a set of QoS parameters. The *traffic contract* is a set of parameters that specify the characteristics of the source traffic. This defines the requirements for compliant cells of the connection. The *QoS parameters* are negotiated by the source with the network, and are used to define the expected quality of service provided by the network. For each service category, the network guarantees the negotiated QoS parameters if the end system complies with the negotiated traffic contract. For non-compliant traffic, the network need not maintain the QoS objective.

The *Constant Bit Rate (CBR)* service category is defined for traffic that requires a constant amount of bandwidth, specified by a Peak Cell Rate (PCR), to be continuously available. The network guarantees that all cells emitted by the source that conform to this PCR will be transferred by the network with minimal cell loss, and within fixed bounds of cell delay and delay variation. The *real time Variable Bit Rate (VBR-rt)* class is characterized by PCR, Sustained Cell Rate (SCR) and a Maximum Burst Size (MBS) in cells that controls the bursty nature of VBR traffic. The network attempts to deliver cells within fixed bounds of cell delay and delay variation. *Non-real-time VBR* sources are also specified by PCR, SCR and MBS, but are less sensitive to delay and delay variation than the real time sources. The network does not specify any delay and delay variation parameters for the VBR-nrt service.

The *Available Bit Rate (ABR)* service category is specified by a PCR and Minimum Cell Rate (MCR) which is guaranteed by the network. The bandwidth allocated by the network to an ABR connection may vary during the life of a connection, but may not be less than MCR. ABR connections use a rate-based closed-loop feedback-control mechanism for congestion control. The network tries to maintain a low Cell Loss Ratio by changing the allowed cell rates (ACR) at which a source can send. The *Unspecified Bit Rate (UBR)* class is intended for best effort applications, and this category does not support any service guarantees. UBR has no built in congestion control mechanisms. The UBR service manages congestion by efficient buffer management policies in the switch. A new service called Guaranteed Frame Rate (GFR) is being introduced at the ATM Forum and the ITU-T. GFR is based on UBR, but guarantees a minimum rate to connections. The service also recognizes AAL5 frames, and performs frame level dropping as opposed to cell level dropping.

In addition, the ITU-T has specified four QoS classes to be used to deliver network based QoS [I35696]. It is imperative that a broadband satellite network be able to support the various QoS services specified by the standards. Most importantly, the network should be able to support TCP/IP based data applications that constitute the bulk of Internet traffic.

Most of the parameters specified in the standards are relevant only to terrestrial networks. These values have to be re-evaluated for Satellite-ATM networks. For example, the ITU-T specifies a maximum cell transfer delay of 400 ms for the ITU Class 1 stringent service [I35696]. This class is expected to carry CBR traffic for real-time voice communications over ATM. However, the 400ms maximum delay needs to be reviewed to ensure that it properly accounts for the propagation delays in geosynchronous satellite networks. The peak-to-peak cell delay variation of QoS Class 1 is also specified to be a maximum of 3 ms by the ITU-T [I35696]. This value may be too stringent for many satellite systems. As a result, the QoS parameters are under careful consideration by ITU-4B [IT4B97] In this context, the ITU-4B preliminary draft recommendations on transmission of Asynchronous Transfer Mode (ATM) Traffic via Satellite is in the process of development.

## 3 Delay Requirements of Applications

We briefly discuss the basic qualitative requirements of three classes of applications, interactive voice/video, non-interactive voice/video and TCP/IP file transfer. Interactive voice requires very low delay (ITU-T specifies a delay of less than 400 ms to prevent echo effects) and delay variation (up to 3 ms specified by ITU-T). GEO systems have a high propagation delay of at least 250 ms from ground terminal to ground terminal. If two GEO hops are involved, then the inter-satellite link delay could be about 240 ms. Other delay components are additionally incurred, and the total end-to-end delay can be higher than 400 ms. Although the propagation and inter-satellite link delays of LEOs are lower, LEO systems exhibit high delay variation due to connection handovers, satellite and orbital dynamics, and adaptive routing. This is further discussed in section 5.3. Non-interactive voice/video applications are real-time applications whose delay requirements are not as stringent as their interactive counterparts. However, these applications also have stringent jitter requirements. As a result, the jitter characteristics of GEO and LEO systems must be carefully studied before they can service real time voice-video applications.

The performance of TCP/IP file transfer applications is throughput dependent and has very loose delay requirements. As a result, both GEOs and LEOs with sufficient throughput can meet the delay requirements of file transfer applications. It is often misconstrued that TCP is throughput limited over GEOs due to the default TCP window size of 64K bytes. The TCP large windows option allows the TCP window to increase beyond 64K bytes and results in the usage of the available capacity even in high bandwidth GEO systems. The efficiency of TCP over GEO systems can be low because the TCP window based flow control mechanism takes several round trips to fully utilize the available capacity. The large round trip time in GEOs results in capacity being wasted during the ramp-up phase. To counter this, the TCP spoof protocol is being designed that splits the TCP control loop into several segments. However this protocol is currently incompatible with end-to-end IP security protocols. Several other mechanisms are being developed to mitigate latency effects over GEOs [GOY97a][TCPS98].

The TCP congestion control algorithm inherently relies on round trip time (RTT) estimates to recover from congestion losses. The TCP RTT estimation algorithm is sensitive to sudden changes in delays as may be experienced in LEO constellations. This may result in false timeouts and retransmits at the TCP layer. More sophisticated RTT measurement techniques are being developed for TCP to counter the effects of delay jitter in LEO systems [TCPS98].

### 3.1 Architectural Issues

Figure 1 illustrates a satellite-ATM network represented by a ground segment, a space segment, and a network control center. The ground segment consists of ATM networks that may be further connected to other legacy networks. The network control center (NCC) performs various management and resource allocation functions for the satellite media. Inter-satellite links (ISL) in the space segment provide seamless global connectivity to the satellite constellation. The network allows the transmission of ATM cells over satellite, multiplexes and demultiplexes ATM cell streams from uplinks and downlinks, and maintains the QoS objectives of the various connection types. The satellite-ATM network also includes a satellite-ATM interface device connecting the ATM network to the satellite system. The interface device transports ATM cells over the frame based satellite network, and demultiplexes ATM cells from the satellite frames. The device typically uses a DAMA technique to obtain media access to the satellite physical layer. The interface unit is also responsible for forward error correction techniques to reduce the error rates of the satellite link. The unit must maintain ATM quality of service parameters at the entrance to the satellite network. As a result, it translates the ATM QoS requirements into corresponding requirements for the satellite network. This interface is thus responsible for resource allocation, error control, and traffic control. Details about this model can be obtained from [KOTA97].

This architectural model presents several design options for the satellite and ground network segments. These options include

1. No on-board processing or switching.

2. On-board processing with ground ATM switching.

3. On-board processing and ATM switching.

About 53% of the planned Ka-band satellite networks propose to use on-board ATM like fast packet switching [PONZ97]. An overview of the network architectures of some of the proposed systems can be found in [WUPI94]. In a simple satellite model without on-board processing or switching, minimal on-board buffering is required. However, if on-board processing is performed, then on-board buffering is needed to achieve the multiplexing gains provided by ATM. On-board processing can be used for resource allocation and media access control (MAC). MAC options include TDMA, FDMA, and CDMA and can use contention based, reservation based, or fixed media access control. Demand Assignment Multiple Access (DAMA) [KOT97b] can be used with any of the MAC options. If on-board processing is not performed, DAMA must be done by the NCC. On-board DAMA decreases the response time of the media access policy by half because link access requests need not travel to the NCC on the ground any more. In addition to media access control, ABR explicit rate allocation or EFCI control, and UBR/GFR buffer management can also be performed on-board the satellite. On-board switching may be used for efficient use of the network by implementing adaptive routing/switching algorithms. Trade-offs must be made with respect to the complexity, power and weight requirements for providing on-board buffering, switching and processing features to the satellite network. In addition, on-board buffering and switching will introduce some additional delays within the space segment of the network. For fast packet or cell switched satellite networks, the switching delay is negligible compared to the propagation delay, but the buffering delay can be significant. Buffering also results in delay variations due to the bursty nature of ATM traffic.

The major focus of this paper includes:

1. The development of an end-to-end satellite network delay model.

2. Simulation and analysis of the buffering requirements of the satellite network for TCP/IP traffic over the UBR service category

In this paper, we have assumed that all processing is performed at the ground terminals with the help of the NCC. The simulations of buffer requirements estimate the buffers needed at the ground stations, and assume that no on-board processing or buffering is performed. However, the delay model presented in the next section is applicable for on-board processing and switching systems.

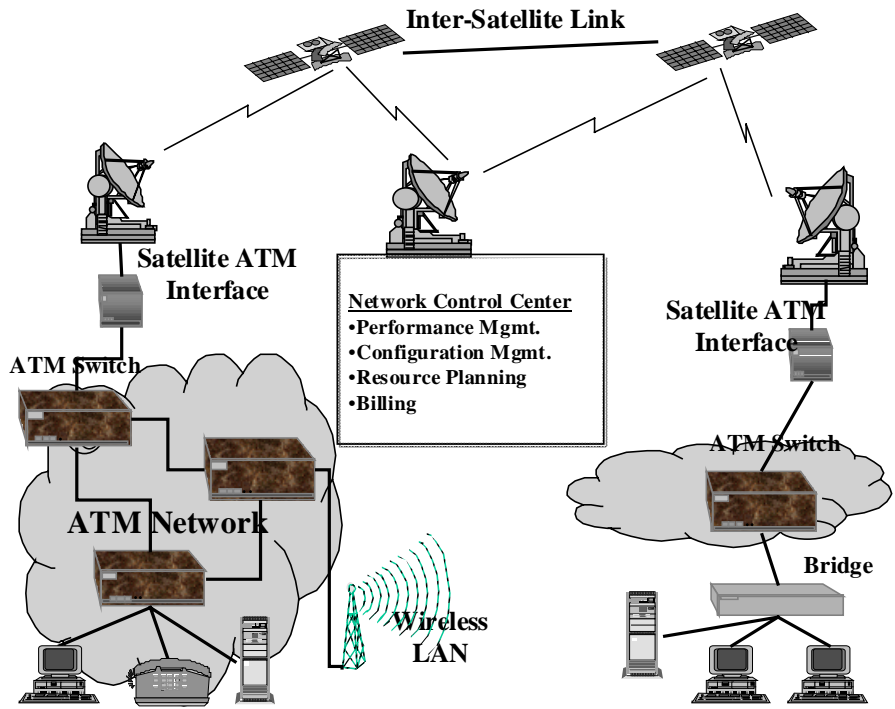

**Figure 1: Satellite-ATM network model**

## 4 Satellite Network Delay Model

In this section, we develop a simple delay model of a satellite network. This model can be used to estimate the end-to-end delay of both GEO and LEO satellite networks.

The end-to-end delay (D) experienced by a data packet traversing the satellite network is the sum of the transmission delay ($t_t$), the uplink ($t_{up}$) and downlink ($t_{down}$) ground segment to satellite propagation delays, the inter-satellite link delay ($t_i$), the on-board switching and processing delay ($t_s$) and the buffering delay ($t_q$). The inter-satellite, on-board switching, processing and buffering delays are cumulative over the path traversed by a connection. In this model, we only consider the satellite component of the delay. The total delay experienced by a packet is the sum of the delays of the satellite and the terrestrial networks. This model does not incorporate the delay variation experienced by the cells of a connection. The delay variation is caused by orbital dynamics, buffering, adaptive routing (in LEOs) and on-board processing. Quantitative analysis of delay jitter in satellite systems is beyond the scope of this study. The end-to-end delay (D) is given by:

$$D = t_t + t_{up} + t_i + t_{down} + t_s + t_q$$

**Transmission delay:** The transmission delay ($t_t$) is the time taken to transmit a single data packet at the network data rate.

$$t_t = \frac{packet\_size}{data\_rate}$$

For broadband networks with high data rates, the transmission delays are negligible in comparison to the satellite propagation delays. For example, a 9180 byte TCP packet is transmitted in about 472 microseconds. This delay is much less than the propagation delays in satellites.

**Propagation delay:** The propagation delay for the cells of a connection is the sum of the following three quantities:

1. The source ground terminal to source satellite propagation delay ($t_{up}$)

2. The Inter-satellite link propagation delays ($t_i$)

3. The destination satellite to destination ground terminal propagation delay ($t_{down}$)

The *uplink and downlink satellite-ground terminal propagation delays* ($t_{up}$ and $t_{down}$ respectively) represent the time taken for the signal to travel from the source ground terminal to the first satellite in the network ($t_{up}$), and the time for the signal to reach the destination ground terminal from the last satellite in the network ($t_{down}$).

$$t_{up} = \frac{source\_satellite\_dist}{speed\_of\_signal}$$
$$t_{down} = \frac{dest\_satellite\_dist}{speed\_of\_signal}$$

The *inter-satellite link delay* ($t_i$) is the sum of the propagation delays of the inter-satellite links (ISLs) traversed by the connection. Inter-satellite links (crosslinks) may be *in-plane* or *cross-plane* links. In-plane links connect satellites within the same orbit plane, while cross-plane links connect satellites in different orbit planes. In GEO systems, ISL delays can be assumed to be constant over a connection's lifetime because GEO satellites are almost stationary over a given point on the earth, and with respect to one another. In LEO constellations, the ISL delays depend on the orbital radius, the number of satellites-per-orbit, and the inter-orbital distance (or the number of orbits). Also, the ISL delays change over the life of a connection due to satellite movement and adaptive routing techniques in LEOs. As a result, LEO systems can exhibit a high variation in ISL delay.

$$t_i = \frac{\sum ISL\_lengths}{speed\_of\_signal}$$

**Buffering delay:** Buffering delay ($t_q$) is the sum of the delays that occur at each hop in the network due to cell queuing. Cells may be queued due to the bursty nature of traffic, congestion at the queuing points (earth stations and satellites), or due to media access control delays. Buffering delays depend on the congestion level, queuing and scheduling policies, connection priority and ATM service category. CBR and real time VBR connections suffer minimum buffering delays because they receive higher priority than the non-real time connections. Cells from ABR and UBR connections could suffer significant delay at each satellite hop during periods of congestion.

**Switching and processing delays:** The data packets may incur additional delays ($t_s$) at each satellite hop depending on the amount of on-board switching and processing. For high data rate networks with packet/cell switching, switching and processing delays are negligible compared to the propagation delays.

## 5 Propagation Delay Model

In this section, we present a propagation delay model for satellite networks. The GEO model is fairly simple due to the stationary nature of GEO satellites, and the small number of satellites needed to cover the earth. The LEO model assumes a circular multi-orbit constellation with evenly spaced orbits and evenly spaced satellites within the orbits.

## 5.1 GEO Propagation Delay Model

GEO systems are at an altitude of about 36,000 km above the equator. For GEOs, $t_{up}$ and $t_{down}$ can be approximated to about 125 ms each for ground terminals near the equator. Inter satellite propagation delays are stable and depend on the number of satellites in the constellation. As few as three GEOs are sufficient to cover the earth. Table 1 lists the inter-satellite link distances and propagation delays for GEO systems with N satellites evenly spaced around the equatorial plane. For ground terminals farther away from the equator, the propagation delay from ground station to ground station through a single satellite is about 275 ms.

**Table 1 : GEO Inter Satellite Delays**

| Number of Satellites (N) | Inter-Satellite Link Distance (km) | Inter-Satellite Link Delay (ms) |
|---|---|---|
| 3 | 73,030 | 243 |
| 4 | 59,629 | 199 |
| 5 | 49,567 | 165 |
| 6 | 42,164 | 141 |
| 7 | 36,589 | 122 |
| 8 | 32,271 | 108 |
| 9 | 28,842 | 96 |
| 10 | 26,059 | 87 |
| 11 | 23,758 | 79 |
| 12 | 21,826 | 73 |

## 5.2 LEO Propagation Delay Model

In this section, we provide a simple model for the propagation delays of LEO systems. This model calculates the total propagation delay from source ground terminal to the destination ground terminal, through the LEO network. A LEO geometry and network topology model is used to determine the total number of satellites and the total propagation delay for communication between two ground points. The model uses the following information.

- Number of orbit planes
- Number of satellites per orbit plane
- Satellite altitude
- Orbit plane inclination angle[1]
- Ground terminal coordinates

### 5.2.1 LEO Orbital Model

Low Earth Orbit (LEO) communication satellites are arranged in a constellation in the following manner. The satellites are organized into a set of *number_of_orbit_planes* orbit planes, each containing *number_of_sats_per_plane* satellites. The orbits are assumed to be circular and to have a common, high inclination angle *(inclination)*. The inclination angle, combined with the electronic-horizon reach of the satellites directly determines the range of latitudes for which the system can provide service. The satellites within a given orbit plane are evenly spaced by using a delta anomaly between in-plane satellite orbits:

$$delta\_anomaly = \frac{360°}{number\_of\_sats\_per\_plane}$$

The orbit planes are approximately evenly spaced about the earth polar axis by using a delta right ascension and a correction term between orbit planes:

$$delta\_right\_ascension = \frac{180°}{number\_of\_orbit\_planes} + RA\_correction$$

Spreading the right ascension of the orbit planes over 180 degrees means that the satellites in adjacent planes are roughly traveling in parallel, with the exception that between the last and first planes, the satellites are traveling in opposite directions. The interval between the last and first orbit plane is called the "seam" of the constellation. The fact that the satellites in the last and first orbit planes of the constellation are travelling in opposite directions means that any cross-plane links spanning the seam will have to change connectivity to a different satellite every few minutes. This will result in frequent handovers causing additional delays and delay variations.

---

[1] Inclination angle is the angle made by the satellite orbital plane with the equatorial plane.

The *RA_correction* term in the previous formula is necessary for the following reason. LEO communication constellations use inclination angles of less than 90 degrees. Because of this, the last orbit plane tends to tilt in the opposite direction as the first orbit plane by roughly twice the complement of the inclination angle (i.e., 2×(90-*inclination*)). Without the correction term for the *delta_right_ascension*, a "hole" results in the ground coverage of the constellation in the two areas of the earth where the seam orbit-planes are tilting away from each other. In the opposite hemisphere of the two holes, the seam orbit-planes are tilting towards each other, resulting in overlapping, or redundant, ground coverage. Trade-offs can be made between how much of the serviced latitude range will be provided continuous, uninterrupted service, and the amount of redundant coverage suffered. The model described here currently uses the following simple correction term.

$$RA\_correction = \frac{1.5 \times (90° - inclination)}{number\_of\_orbit\_planes}$$

The inter-plane satellites are phased by about one-half of the in-plane satellite spacing. This staggered phasing provides a more optimal and uniform coverage pattern, and maximizes inter-plane satellite distances near the extreme latitudes where the orbit planes cross. The current model uses the following delta inter-plane phasing.

$$Delta\_inter\_plane\_phasing = 0.5 \times delta\_anomaly + delta\_right\_ascension \times \sin(90 - inclination)$$

The model uses an Earth Centered Inertial (ECI) right-handed coordinate system. The first and second axes lie on the equatorial plane with the first axis aligned with the prime meridian. The third axis is aligned with the earth's polar axis pointing north. The first satellite of the first orbit plane (satellite(1,1)) is arbitrarily placed at latitude, longitude coordinates of (0,0), giving this satellite a position vector in ECI coordinates of ($r$, 0, 0), where $r$ is the sum of the equatorial earth radius and the satellite altitude. The position vectors of the remaining satellites are obtained by an appropriate series of rotations of the satellite(1,1) position vector involving the angles described above.

**5.2.2 LEO Route Calculation**

We use this model to calculate routes and propagation delays across the satellite constellation. We first use the above procedure to create a constellation pattern providing continuous ground coverage for most of the latitude range covered by the constellation. Circular orbits with staggered inter-plane phasing are assumed as described above. Each satellite is assumed to have four crosslinks (inter-satellite links) providing connectivity to the in-plane satellites immediately leading and following, and to the nearest satellites in the two adjacent orbit planes -- in navigational terms, these would be the fore, aft, port, and starboard satellites. Cross-plane connectivity constraints in the area of extreme latitudes or across the constellation seam (i.e., where satellites in the last orbit plane and the first orbit plane are traveling in opposite relative directions) are not considered. Anticipatory routing to reduce hand-offs and routing around congested paths is not considered.

The following simple algorithm is used to determine the route between two ground points through the satellite constellation. One of the ground points is designated as the source node and the other ground point is assigned as the destination node. The satellite nearest to the source node, and the satellite nearest to the destination node are first determined. This assumes minimal redundant satellite ground coverage, which is usually the case for LEO communication systems. Starting at the satellite nearest the source node, of the satellites with which it has connectivity, the satellite nearest to the destination node's satellite is selected.  The process is repeatedly applied at each selected satellite, with backtracking precluded, until the destination node's satellite is reached. The algorithm then counts the number of satellites in the end-to-end, source-to-destination path. The distances between successive path-nodes, beginning at the source terminal and ending at the destination terminal, are computed, converted to link propagation delays by dividing by the speed of light, and accumulated to provide the path end-to-end propagation delay.  The number of satellites in the route path and the total propagation delay are then reported. While the routing algorithm just described is strictly geometry based and only locally optimal, of the limited cases examined thus far, the results appear to be generally coincident with a globally optimal solution.

The model can also generate three-dimensional orthographic-projection displays showing satellite orbits, satellite positions, cross-links, ground terminal positions, path-links followed, and earth model from any desired viewing direction. Figure 2 shows an example path of a 6-plane, 11-satellites per plane LEO system path from Los Angeles to London. The associated configuration parameters are given in Table 2. Table 3 shows the resultant number of path-satellites, the individual link delays, and the total end-to-end delay for the example. Table 4 shows the end-to-end propagation delays for a 6-plane, 11-satellites per plane constellation, between 10 cities of the world ranked by Gross Domestic Product. Table 5 shows the number of satellites in the path between the same set of cities.  Table 6 and Table 7 show the same information for a 12-plane, 24-satellites per plane constellation at an altitude of 1400 km and an inclination angle of 82 degrees.

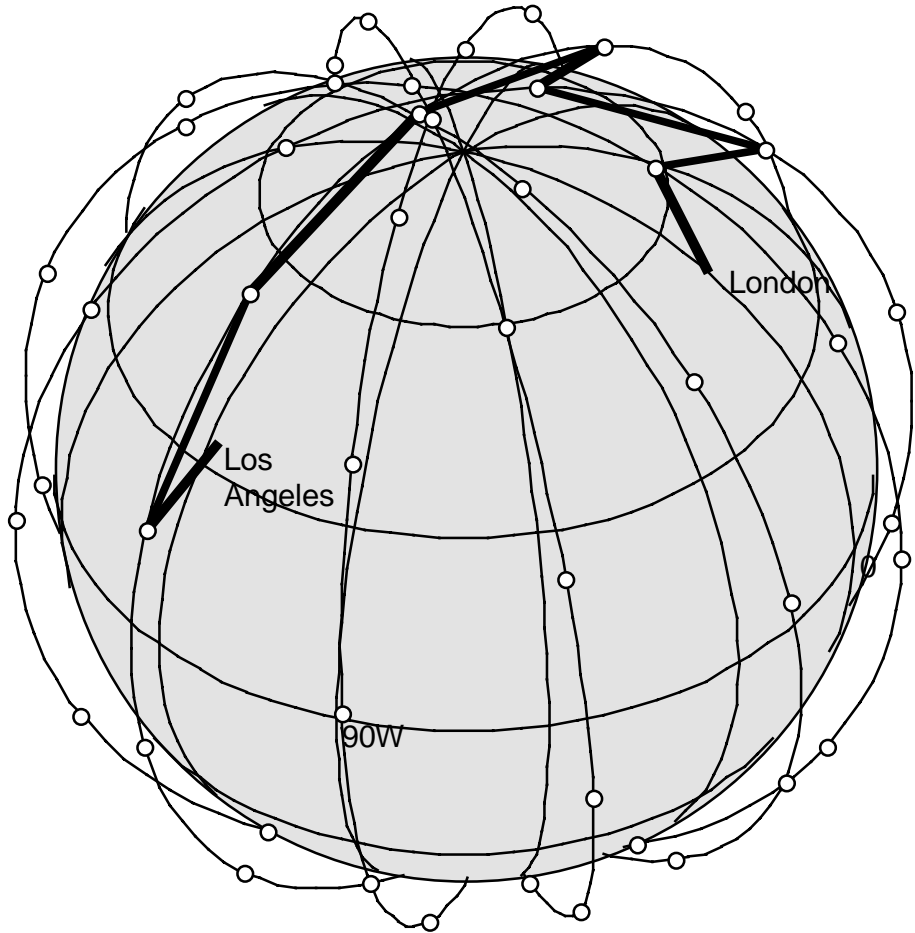

**Figure 2: Example path through LEO constellation**

**Table 2: LEO configuration parameters**

| Orbit Planes | 6 |
|---|---|
| Sats Per Plane | 11 |
| Total Sats | 66 |
| Altitude(km) | 780 |
| Inclination(deg) | 86 |

**Table 3: LEO propagation delays**

| Source | Los Angeles |
|---|---|
| Destination | London |
| Satellites In Path | 7 |
| **Propagation Delays(ms)** | |
| Uplink | 5.87 |
| Downlink | 6.37 |
| ISL 1 | 13.44 |
| ISL 2 | 13.44 |
| ISL 3 | 13.44 |
| ISL 4 | 7.80 |
| ISL 5 | 13.44 |
| ISL 6 | 9.63 |
| **Total Prop. Delay** | **83.45** |

**Table 4: Total propagation delays in milliseconds in city-to-city path for 6x11 constellation**

|  | New York | Tokyo | Paris | London | Seoul | Los Angeles | Toronto | Mexico City | Sydney | Chicago |
|---|---|---|---|---|---|---|---|---|---|---|
| New York | 11 | | | | | | | | | |
| Tokyo | 58 | 13 | | | | | | | | |
| Paris | 60 | 60 | 13 | | | | | | | |
| London | 39 | 47 | 26 | 13 | | | | | | |
| Seoul | 62 | 37 | 83 | 68 | 11 | | | | | |
| Los Angeles | 24 | 71 | 69 | 83 | 71 | 12 | | | | |
| Toronto | 10 | 57 | 54 | 38 | 63 | 23 | 9 | | | |
| Mexico City | 26 | 73 | 51 | 61 | 79 | 39 | 25 | 14 | | |
| Sydney | 62 | 37 | 91 | 71 | 36 | 49 | 61 | 77 | 7 | |
| Chicago | 8 | 56 | 53 | 36 | 61 | 22 | 7 | 23 | 59 | 6 |

**Table 5: Number of satellites in city-to-city path for 6x11 constellation**

|  | New York | Tokyo | Paris | London | Seoul | Los Angeles | Toronto | Mexico City | Sydney | Chicago |
|---|---|---|---|---|---|---|---|---|---|---|
| New York | 1 | | | | | | | | | |
| Tokyo | 5 | 1 | | | | | | | | |
| Paris | 5 | 5 | 1 | | | | | | | |
| London | 4 | 4 | 2 | 1 | | | | | | |
| Seoul | 5 | 3 | 7 | 6 | 1 | | | | | |
| Los Angeles | 2 | 6 | 6 | 7 | 6 | 1 | | | | |
| Toronto | 1 | 5 | 5 | 4 | 5 | 2 | 1 | | | |
| Mexico City | 2 | 6 | 4 | 5 | 6 | 3 | 2 | 1 | | |
| Sydney | 5 | 3 | 7 | 6 | 3 | 4 | 5 | 6 | 1 | |
| Chicago | 1 | 5 | 5 | 4 | 5 | 2 | 1 | 2 | 5 | 1 |

**Table 6: Total propagation delays in milliseconds in city-to-city path for 12x24 constellation.**

|  | New York | Tokyo | Paris | London | Seoul | Los Angeles | Toronto | Mexico City | Sydney | Chicago |
|---|---|---|---|---|---|---|---|---|---|---|
| New York | 10 | | | | | | | | | |
| Tokyo | 57 | 11 | | | | | | | | |
| Paris | 77 | 109 | 10 | | | | | | | |
| London | 78 | 110 | 11 | 12 | | | | | | |
| Seoul | 58 | 24 | 75 | 76 | 11 | | | | | |
| Los Angeles | 41 | 57 | 99 | 100 | 59 | 10 | | | | |
| Toronto | 10 | 57 | 73 | 74 | 58 | 44 | 11 | | | |
| Mexico City | 24 | 67 | 87 | 88 | 68 | 30 | 24 | 10 | | |
| Sydney | 92 | 52 | 89 | 89 | 53 | 115 | 92 | 113 | 10 | |
| Chicago | 22 | 54 | 83 | 84 | 57 | 30 | 23 | 23 | 101 | 10 |

Table 7: Number of satellites in city-to-city path for 12x24 constellation

|             | New York | Tokyo | Paris | London | Seoul | Los Angeles | Toronto | Mexico City | Sydney | Chicago |
|-------------|----------|-------|-------|--------|-------|-------------|---------|-------------|--------|---------|
| New York    | 1        |       |       |        |       |             |         |             |        |         |
| Tokyo       | 8        | 1     |       |        |       |             |         |             |        |         |
| Paris       | 10       | 15    | 1     |        |       |             |         |             |        |         |
| London      | 10       | 15    | 1     | 1      |       |             |         |             |        |         |
| Seoul       | 8        | 3     | 12    | 12     | 1     |             |         |             |        |         |
| Los Angeles | 6        | 9     | 15    | 15     | 9     | 1           |         |             |        |         |
| Toronto     | 1        | 8     | 10    | 10     | 8     | 6           | 1       |             |        |         |
| Mexico City | 3        | 10    | 12    | 12     | 10    | 4           | 3       | 1           |        |         |
| Sydney      | 14       | 7     | 12    | 12     | 7     | 19          | 14      | 18          | 1      |         |
| Chicago     | 3        | 8     | 12    | 12     | 8     | 4           | 3       | 3           | 13     | 1       |

## 5.3 Delay Variation Characteristics

Although LEO networks have relatively smaller propagation delays than GEO networks, the delay variation in LEOs can be significant. The delay variation in LEO systems can arise from several factors:

1. **Handovers:** The revolution of the satellites within their orbits causes them to change position with respect to the ground terminals. As a result, the ground terminal must handover the connections from the satellite descending below the horizon to the satellite ascending from the opposing horizon. Based on the velocity, altitude and the coverage of the satellites, it is estimated that call handovers can occur on an average of every 8 to 11 minutes [IQTC97]. The handover procedure requires a state transfer from one satellite to the next, and will result in a change in the delay characteristic of the connection at least for a short time interval. If the satellites across the seam of the constellation are communicating via crosslinks, the handover rate is much more frequent because the satellites are travelling in opposite directions.

2. **Satellite Motion:** Not only do the satellites move with respect to the ground terminal, they also move relative to each other. When satellites in adjacent orbits cross each other at the poles, they are now traveling in opposite sides of each other. As a result, calls may have to be rerouted accordingly resulting in further changes in delays.

3. **Buffering and Processing:** A typical connection over a LEO system might pass through several satellites, suffering buffering and processing delays at each hop. For CBR traffic, the buffering delays are small, but for bursty traffic over real time VBR (used by video applications), the cumulative effects of the delays and delay variations could be large depending on the burstiness and the amount of overbooking in the network.

4. **Adaptive Routing:** Due to the satellite orbital dynamics and the changing delays, most LEO systems are expected to use some form of adaptive routing to provide end-to-end connectivity. Adaptive routing inherently introduces complexity and delay variation. In addition, adaptive routing may result in packet reordering. These out of order packets will have to be buffered at the edge of the network resulting in further delay and jitter.

GEO systems exhibit relatively stable delay characteristics because they are almost stationary with respect to the ground terminals. Connection handovers are rare in GEO systems and are mainly due to fault recovery reasons. As a result, there is a clear trade-off between delay and jitter characteristics of GEO and LEO systems, especially for interactive real-time applications.

## 6 Buffering Delays for TCP/IP over Satellite-ATM ABR and UBR Service Classes

The majority of Internet traffic is TCP/IP based data traffic. It is thus important to assess the buffering characteristics for TCP/IP applications over satellite-ATM networks. Most TCP/IP data applications will use the ABR or UBR service categories in ATM networks. The maximum buffering delay can be calculated from an estimate of the buffer size at each queuing point in the connection's path:

$$Buffering\_delay \leq \frac{Buffer\_size}{Buffer\_drain\_rate}$$

The buffer drain rate is the rate at which cells are serviced from the buffer. This rate depends on the link capacity, the scheduling policy, and other higher priority traffic on the link. The queuing points in the network must have sufficient buffer size to ensure good performance of TCP/IP applications. In this section, we present a simulation model to calculate the buffer requirements for TCP/IP traffic over ATM networks for the ABR and UBR service categories. Section 6.1 outlines some known results on buffer requirements of TCP over ABR [SHIV98]. Section 6.2 describes a simulation model for TCP/IP over satellite-ATM UBR, and presents simulation results to estimate the buffer requirements for TCP/IP file transfer traffic. The estimates of buffer requirements are then used to calculate the queuing delays at each queuing point in the connection's path.

## 6.1 Buffer Requirements for TCP/IP over ABR

An analysis of the buffer requirements of TCP/IP over ABR has been conducted in [SHIV97] and [SHIV98]. ABR uses a rate based, closed loop feedback control model for congestion control. As a result, the performance of the ABR service depends on the ABR congestion control scheme used in the switch, and on the ABR source-end-system parameters. In general, a good ABR switch scheme should be able to control queues within the ATM network. The ERICA+ (Explicit Rate Indication Congestion Avoidance +) scheme [ERIC97] has been specified as a sample scheme by the ATM Forum. [SHIV97] and [SHIV98] show that for the ERICA+ scheme, the buffer requirements for an ABR switch to ensure zero packet loss for TCP/IP traffic over ABR can be bounded by a constant multiple of the round trip propagation delay from the ABR end-system or virtual end-system to the bottleneck ABR node in the network. The delay is called the *feedback delay* of the network, and it specifies the time taken for the effect of the bottleneck feedback to be seen by the network. The feedback delay at a queuing point can be restricted to the round trip propagation delay from the previous end-system by implementing Backward Explicit Congestion Notification (BECN) or Virtual Source/Virtual Destination (VS/VD) at the satellite nodes. For TCP/IP file transfer traffic, the buffer requirements are proportional only to the feedback delay, and are independent of the number of TCP sources and other background traffic in the ATM network. Thus, TCP connections can be transported through a finite buffer ABR network with zero packet loss.

## 6.2 Buffer Requirements for TCP/IP over UBR

Most ATM networks are expected to be implemented as backbone networks within an IP based Internet where edge devices separate ATM networks from IP networks. Currently, IP networks do not support the rate based flow control mechanisms used by ABR. The above studies have shown that for ATM in the backbone, the buffer requirements of nodes at the periphery of the ATM network (edge devices) for TCP/IP traffic are comparable to buffer requirements for TCP/IP with UBR. Moreover, since TCP has its own flow and congestion control mechanisms, many TCP/IP connections are expected to use the UBR service. As a result, it is important to assess the buffer sizes (and hence delays) at UBR queuing points in a satellite network.

In this subsection, we present a simulation model for calculating the buffer requirements for satellite-UBR networks to efficiently support TCP/IP traffic. We present results of SACK TCP throughput over satellite-UBR for various satellite latencies, buffer sizes and number of sources.

### 6.2.1 Simulation Model

Figure 3 shows the basic network configuration used in the paper to assess buffer requirements at a single bottleneck node. In the figure, the switches represent the earth stations that connect to the satellite constellation. The earth stations interface the terrestrial network with the satellite network. In general, the satellite network model may include on-board processing and queuing. In the results stated in this section, no on-board processing or queuing is performed. The bottleneck node is the earth station at the entry to the satellite network. As a result, in our experiments, no queuing delays occur in the satellite network. All processing and queuing are performed at the earth stations. The goal of this study is to assess the buffer requirements of the bottleneck node (in this case, the earth station) for good TCP/IP performance.

All simulations use the N source configuration shown in the figure. All sources are identical and persistent TCP sources. The TCP layer always sends a segment as long as it is permitted by the TCP window. Moreover, traffic is unidirectional so that only the sources send data. The destinations only send ACKs. The TCP delayed acknowledgement timer is deactivated, and the receiver sends an ACK as soon as it receives a segment. TCP with selective acknowledgments (SACK TCP) is used in our simulations. All link bandwidths are 155.52 Mbps, and peak cell rate at the ATM layer is 149.7 Mbps. This accounts for a SONET like overhead in the satellite component of the network.

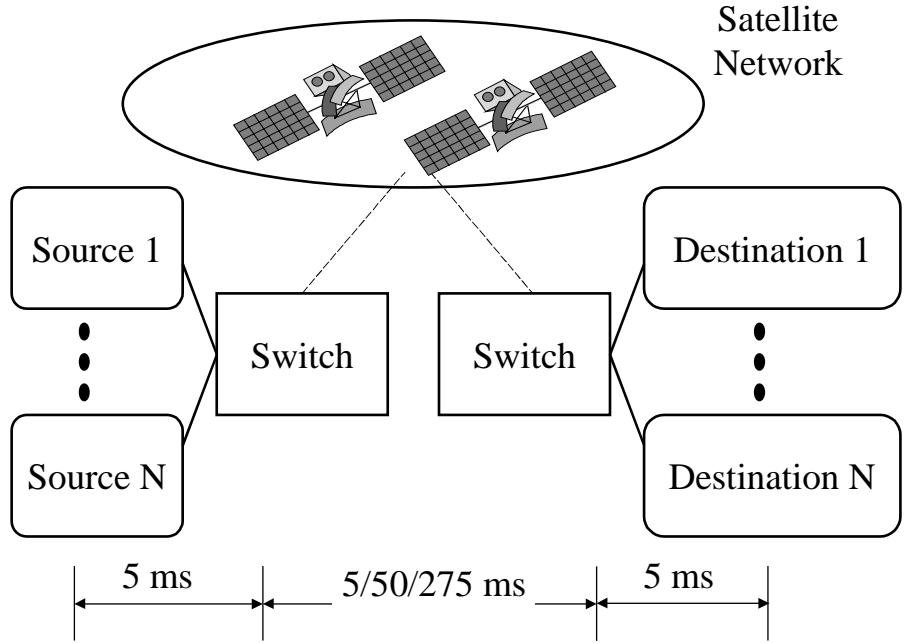

**Figure 3: Simulation model for TCP/IP over UBR**

The following parameters are used to assess the buffer requirements:

**Latency:** Our primary aim is to study the buffer requirements for long latency connections. A typical latency from earth station to earth station for a single LEO hop is about 5 ms. The latencies for multiple LEO hops can easily be 50 ms or more from earth station to earth station. GEO latencies are typically 275 ms from earth station to earth station for earth stations that are not on the equator. We study these three latencies (5 ms, 50 ms, and 275 ms) with various number of sources and buffer sizes. The link delays between the switches and the end systems are 5 ms in all configurations. This results in round trip propagation delays (RTT) of 30 ms, 120 ms and 570 ms respectively.

**Number of sources:** To ensure that the recommendations are scalable and general with respect to the number of connections, we will use configurations with 5, 15 and 50 TCP connections on a single bottleneck link. For single hop LEO configurations, we use 15, 50 and 100 sources.

**Buffer size:** This is the most important parameter of this study. The goal is to estimate the smallest buffer size that results in good TCP performance, and is scalable to the number of TCP sources. The values chosen for the buffer size are approximately:

$$Buffer\_size = 2^{-k} \times RTT \times bottleneck\_link\_data\_rate, k = -1..6$$

i.e., we choose 2, 1, 0.5, 0.25, 0.125, 0.0625, 0.031 and 0.016 multiples of the round trip delay-bandwidth product of the TCP connections. The resulting buffer sizes (in cells) used in the earth stations are as follows:
- *Single LEO:* 375, 750, 1500, 3000, 6000, 12000 (=1 RTT), 24000 and 36000 cells.
- *Multiple LEO:* 780, 1560, 3125, 6250, 12500, 25000, 50000 (=1 RTT), and 100000 cells.
- *GEO:* 3125, 6250, 12500, 25000, 50000, 100000, 200000 (=1 RTT), and 400000 cells.

We plot the buffer size against the achieved TCP throughput for different delay-bandwidth products and number of sources. The asymptotic nature of this graph provides information about the optimal buffer size for the best performance.

**Buffer allocation policy:** We use a per-VC buffer allocation policy called *selective drop* [GOY97a][STAL98]to fairly allocate switch buffers to the competing TCP connections.

**End system policies:** We use an enhanced version of TCP called SACK TCP [RF2018], for this study. SACK TCP improves performance by using selective acknowledgements for retransmission. Further details about our SACK TCP implementation can be found in [GOY97a]. The maximum value of the TCP receiver window is 600000 bytes, 2500000 bytes and 8704000 bytes for single hop LEO, multiple hop LEO and GEO respectively. These window sizes are obtained using the TCP window scaling option, and are sufficient to achieve full utilization on the 155.52 Mbps links. The TCP maximum segment size is 9180 bytes. This conforms to the segment size recommended for TCP connections over long latency connections. The TCP timer granularity is set to 100 ms. This value limits the time taken for retransmissions to multiples of 100 ms. The value is chosen to balance the attainable throughput with the limitations of the TCP RTT measurement algorithm. With large granularity, TCP could wait a long time before detecting packet loss, resulting in poor throughput. Finer granularity of the retransmission timer leads to false timeouts even with a small variation in the measured RTT values.

**6.2.2 Performance Metrics**

The performance of TCP over UBR is measured by the *efficiency* and *fairness* which are defined as follows:

$$Efficiency = \frac{\sum_{i=1}^{N} x_i}{x_{max}}$$

Where $x_i$ is the throughput of the $i^{th}$ TCP connection, $x_{max}$ is the maximum TCP throughput achievable on the given network, and $N$ is the number of TCP connections. The TCP throughputs are measured at the destination TCP layers. Throughput is defined as the total number of bytes delivered to the destination application, divided by the total simulation time. The results are reported in Mbps. The maximum possible TCP throughput ($x_{max}$) is the throughput attainable by the TCP layer running over UBR on a 155.52 Mbps link. For 9180 bytes of data (TCP maximum segment size), the ATM layer receives 9180 bytes of data + 20 bytes of TCP header + 20 bytes of IP header + 8 bytes of LLC header + 8 bytes of AAL5 trailer. These are padded to produce 193 ATM cells. Thus, each TCP segment results in 10229 bytes at the ATM layer. From this, the maximum possible throughput = 9180/10229 = 89.7% = 135 Mbps approximately on a 155.52 Mbps link.

$$Fairness = \frac{\sum_{i=1}^{N}\left(\frac{x_i}{e_i}\right)^2}{N \times \left(\sum_{i=1}^{N}\frac{x_i}{e_i}\right)^2}$$

Where $e_i$ is the expected throughput of the $i^{th}$ TCP connection. Both metrics lie between 0 and 1, and the desired values of efficiency and fairness are close to 1 [JAIN91]. In the symmetrical configuration presented above,

$$e_i = \frac{x_{max}}{N}$$

and the fairness metric represents a equal share of the available data rate. For more complex configurations, the fairness metric specifies max-min fairness [JAIN91].

### 6.2.3 Simulation Results

Figures 4, 5, and 6 show the resulting TCP efficiencies for the 3 different latencies. Each point in the figure shows the efficiency (total achieved TCP throughput divided by maximum possible throughput) against the buffer size used. Each figure plots a different latency, and each set of points (connected by a line) in a figure represents a particular value of N (the number of sources).

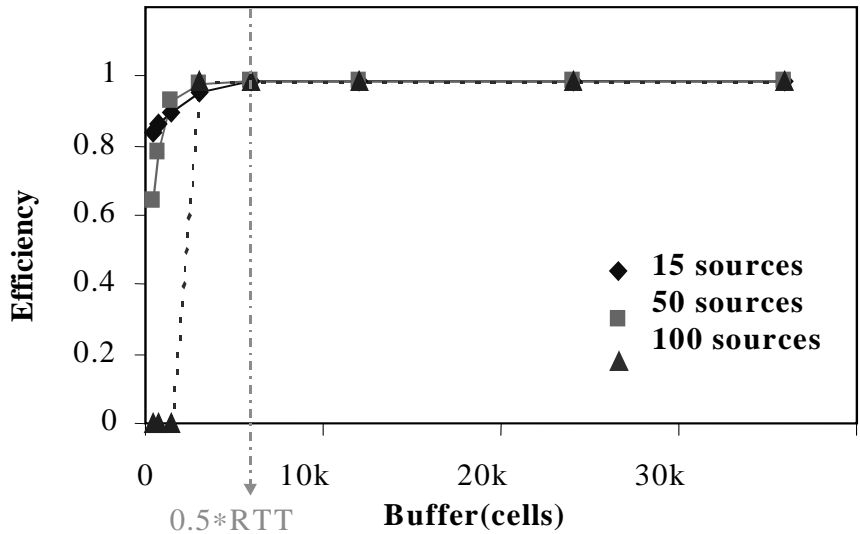

**Figure 4: TCP/IP UBR buffer requirements for single hop LEO**

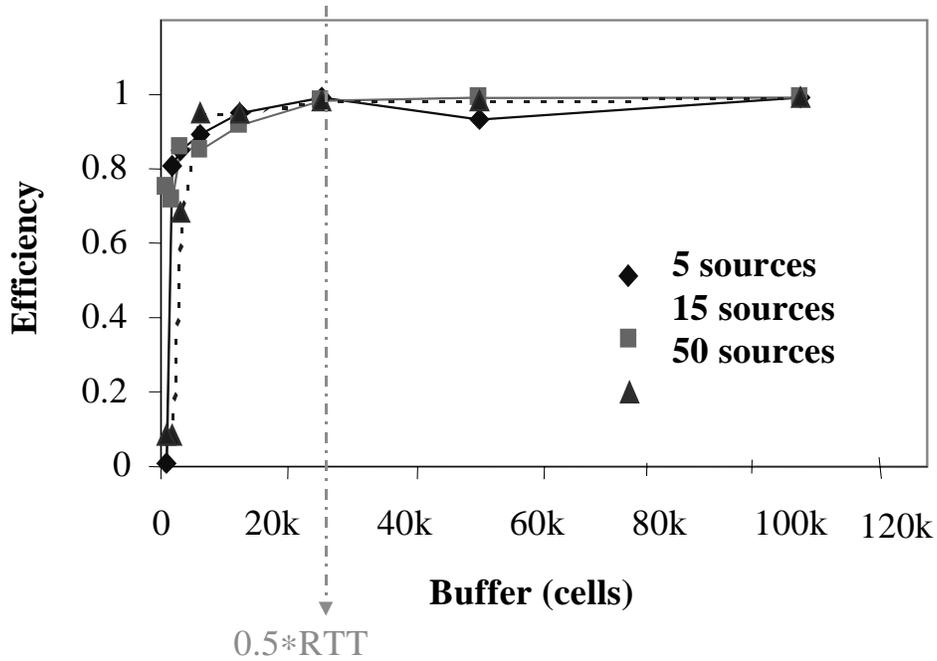

**Figure 5: TCP/IP UBR buffer requirements for multiple hop LEO**

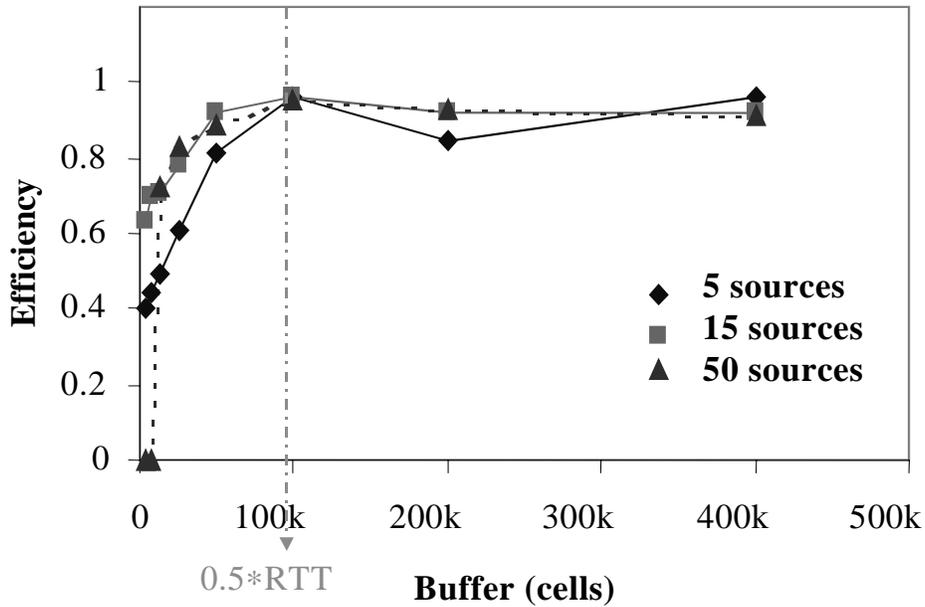

**Figure 6: TCP/IP UBR buffer requirements for single hop GEO**

The following conclusions can be drawn from the figures:

For very small buffer sizes, (0.016RTT, 0.031RTT, 0.0625RTT), the resulting TCP throughput is very low. In fact, for a large number of sources (N=50), the throughput is sometimes close to zero. For small buffer sizes, the performance of TCP/IP deteriorates with increasing number of sources. This is because more TCP packets are dropped from each connection causing TCP timeout and retransmissions. This results in decreased throughput.

For moderate buffer sizes (less then 1 round trip delay times bandwidth), TCP throughput increases with increasing buffer sizes. TCP throughput asymptotically approaches the maximal value with further increase in buffer sizes.

TCP performance over UBR for sufficiently large buffer sizes is scalable with respect to the number of TCP sources. The throughput is never 100%, but for buffers greater than 0.5xRTT, the average TCP throughput is over 98% irrespective of the number of sources.

The knee of the buffer versus throughput graph is more pronounced for larger number of sources. For a large number of sources, TCP performance is very poor for small buffers, but jumps dramatically with sufficient buffering and then stays about the same. For smaller number of sources, the increase in throughput with increasing buffers is more gradual. With sufficient buffers, TCP dynamics enable the connections to share buffer space and link capacity. TCP windows are controlled by the rate at which ACKs are received by the source. The total amount of unacknowledged data is thus controlled by the connection's bandwidth delay product. All this data can be queued at a single queuing point within the network. As a result, each queuing point must have sufficient buffers to support one delay-bandwidth product worth of TCP data so that it can ensure minimal loss.

For large round trip delays, and a small number of sources, a buffer of 1 RTT or more can result in reduced throughput. This is because of the variability in the connection's measured RTT due to buffering delays. When the queuing delay is of the order of the round trip propagation delay, the retransmission timeout values become highly variable. During the initial phase (startup exponential increase), when the queuing delays are small, the timeout value corresponds to the propagation RTT. When the windows increase to fill the switch buffer, the queuing delay increases to about 1 RTT (for a buffer size of about 1 RTT), and packets at the tail of the queue get dropped. Retransmitted packets are sent out after 3 duplicate ACKS are received. However, these retransmitted packets are queued behind a whole RTT worth of queues at the bottleneck switch. As a result, before the sender gets an ACK for retransmitted packets, a timeout occurs, and slow start is incurred. At this point, the sender starts to retransmit from the last unacknowledged segment, but soon receives an ACK for that segment (because the segment was not really lost, but the delay was incorrectly estimated). The loss in throughput occurs during to the time lost in waiting for the retransmission timeout. With smaller buffers, the variability in the RTT is smaller, and false timeouts do not occur. Also, the negative effects of large buffers is not seen in the single hop LEO configuration, because the RTT in this case is much smaller than the timer granularity. As a result, even a high queuing delay is not enough to exceed the minimum timeout value.

The simulation results show that TCP sources with a good per-VC buffer allocation policy like selective drop, can effectively share the link bandwidth. A buffer size of about 0.5RTT to 1RTT is sufficient to provide over 98% throughput to infinite SACK TCP traffic for long latency networks and a large number of sources. This buffer requirement is independent of the number of sources. The fairness in the throughputs measured by the fairness index is high due to the selective drop policy [KOTA97].

## 7 Delay Analysis: Case Study

In this section, we evaluate the end-to-end delay performances of LEO and GEO systems presented in this paper. We consider end-to-end the delay of a connection from New York to Paris. The connection is serviced by a single GEO satellite. For the 6x11 LEO constellation, the connection passes through 5 satellites. For the 12x24 constellation, the connection passes through 10 satellites. The corresponding propagation delays (up-down plus inter-satellite link) can be found from tables 1, 4 and 6. The one way propagation delays from ground station to ground station are 60 ms, 77 ms and 250 ms for the 6x11 LEO, 12x24 LEO and GEO networks respectively. We assume a TCP/IP application over the UBR service category, similar to the one simulated in the previous section. We consider a single queuing point as before, as well as multiple queuing points possibly on-board the satellites. As discussed in the previous section, each queuing point including the ground terminal could have a buffer size of about 0.5RTT for the connection. The resulting buffer sizes are 60 ms for the 6x11 network, 77 ms for the 12x24 network, and 250 ms for the GEO network. This means that a TCP connection can suffer a delay of 60 ms, 77 ms or 250 ms at each queuing point in the respective networks.

Table 8: New York to Paris: Delay Analysis

| Delay | GEO (ms) 1 satellite | 6x11 LEO (ms) 5 satellites | 12x24 LEO (ms) 10 satellites |
|---|---|---|---|
| Transmission | Negligible | Negligible | Negligible |
| Propagation (up+down+ISL) | 250 | 60 | 77 |
| Switching and Processing | Negligible | Negligible | Negligible |
| Buffering (N queuing points) | 0 to N*250 | 0 to N*60 | 0 to N*77 |
| **Total Delay** | **250 to 500** | **60 to 420** | **77 to 924** |

Table 8 lists the individual and total delays for the connection. The transmission and processing delays on a 155.52 Mbps link are small compared to the propagation delay and is thus ignored. From the table, it can be seen that the minimum delay for GEO systems is large in comparison to LEO systems. However, for TCP/IP over UBR traffic, the maximum delays for LEOs are comparable and even higher for the 12x24 system than the GEO system. Moreover, TCP/IP RTT measurement algorithms might experience large variations in delay in LEO systems.

## 8 Summary

In this paper we presented a model to analyze the delay performance of GEO and LEO systems. We first presented a satellite-ATM network architecture model for QoS guarantees over satellite systems. The architecture presents a trade-off between the on-board switching/processing features and the complexity of the satellite communication systems. The end-to-end delay of a connection passing through a satellite constellation consists of the transmission delay, the uplink and downlink propagation delays, the inter-satellite link propagation delays, the satellite switching and processing delays, and the buffering delays. The uplink and downlink propagation delays are much larger in GEO systems than LEO systems because of the higher altitude of the GEO satellites. However, LEO systems can have high delay variations due to orbital dynamics, and connection handovers. The buffering delay for TCP/IP traffic depends on the buffers at each queuing point in the network. The per-hop buffering delay for TCP/IP over ATM-UBR can be about 0.5RTT of the TCP connection. We presented case studies and calculated end-to-end delays of a sample connection from New York to Paris, and concluded that while GEO systems have a large propagation delay, buffering delay can be significant in both GEO and LEO networks.

We have not presented a quantitave analyis of the delay variation experienced by LEO connections. This analysis will lead to greater insights into the feasibility of using satellite networks to support voice, video and data services. A robust technique is needed to mitigate the effect of long delay paths in TCP connections. Protocols like the TCP spoof protocol, or the ABR virtual source / virtual destination option (VS/VD) need to be studied for their feasibility. Optimal routing algorithms in LEOs are also a topic of further study.